\documentclass[prb,twocolumn,superscriptaddress,showpacs]{revtex4}

\usepackage{bm,amsmath,amssymb,graphicx}

\begin{document}

\title{Tuning electronic properties of corrugated graphene: confinement, curvature and band gap opening}

\author{Victor Atanasov}
\affiliation{SQIG, Instituto de Telecomunica\c{c}\~oes,
Av. Rovisco Pais, P-1049-001 Lisbon,
Portugal}
\email{vatanaso@gmail.com}

\author{Avadh Saxena}
\affiliation{Theoretical Division and Center for Nonlinear Studies,
Los Alamos National Laboratory, Los Alamos, NM 87545 USA}
\email{avadh@lanl.gov}

\begin{abstract}
It is shown that for monolayer graphene electrons are confined on a perfect two dimensional surface. The implications for the electronic properties of corrugated graphene are discussed in view of a derivation of the constrained relativistic dynamics for the massless carriers in two dimensions. Surface curvature is related to a series of phenomena with practical applications such as curvature induced p-n junctions, band gap opening and decoherence.  We also establish a bending free energy by treating graphene as a soft electronic membrane. 
\end{abstract}


\maketitle

\section{Introduction}

\begin{figure}[b]
\begin{center}
\includegraphics[scale=0.4]{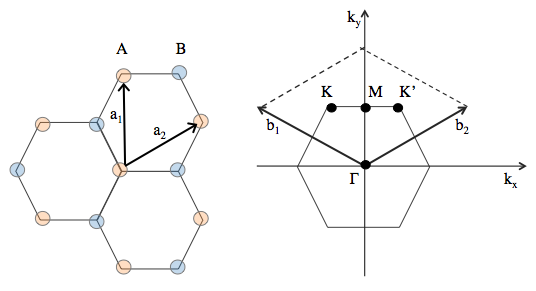}
\caption{\label{graphene_lattice} Left: The lattice structure of graphene is  comprised of two interpenetrating triangular lattices A and B with lattice unit vectors $\rm a_1$ and $\rm a_2.$ Right: The Brillouin zone where the Dirac cones are located, $\rm K$ and $\rm K'.$ The reciprocal lattice vectors are $\rm b_1$ and $\rm b_2.$ }
\end{center}
\end{figure}

\begin{figure}[ht]
\begin{center}
\includegraphics[scale=0.4]{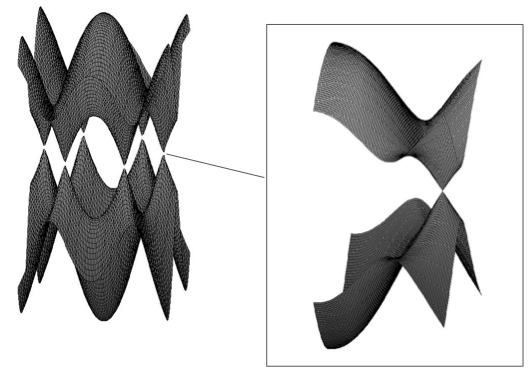}
\caption{\label{graphene_el} Left: The electronic structure of graphene \cite{3} $E_{\pm} = \pm t \sqrt{3+f(\vec{k})} - t' f(\vec{k})$, where $f(\vec{k})=2 \cos{\left(\sqrt{3} k_y a\right)}+4 \cos{\left(\frac{\sqrt{3}}{2} k_y a\right)} \cos{\left(\frac{3}{2} k_x a\right)}.$ Here $\vec{k}=(k_x, k_y),$ $t=2.8$ eV \cite{review}, $t'=0.1$ eV \cite{Deacon} and $a=1.42 $ \AA.  The plus sign applies to the upper ($\pi$) and the minus sign to the lower ($\pi^{\ast}$) band. Right: The band structure zoomed in on one of the Dirac points where $E_{\pm}=\pm \hbar v_F \left|\vec{k}\right|.$ The Fermi velocity $v_F$ is constant and the electron-hole symmetry is not broken if next-to-nearest neighbor hopping is neglected.  }
\end{center}
\end{figure}

The two dimensional form of $\rm sp^2$ hybridized carbon, graphene, is a flexible one-atom-thick soft membrane embedded in three dimensional space. Due to an anharmonic (nonlinear) coupling between bending and stretching phonon modes \cite{1}, it comes with ripples whose typical height scales with sample size $L$ as $L^t$ , with $t=0.6.$ Besides providing structural stability, ripples also play an important role in graphene's electronic properties \cite{2}. Graphene has two atoms per unit cell, which results in two ÔconicalÕ points  $\rm K$ and $\rm K'$ per Brillouin zone where band crossing occurs \cite{5}, see Fig. \ref{graphene_lattice}. Near these crossing points, the electron energy is linearly dependent on the wave vector, see Fig. \ref{graphene_el}. This behavior is robust with respect to long-range hopping processes beyond simple nearest-neighbor, tight-binding approximation \cite{3} because it follows from symmetry considerations \cite{4}. Here we consider the constrained relativistic problem for the carriers in two dimensions and discuss the effects that arise from the emergence of a geometry induced potential $V_G$ on the electronic properties.

The effect of random surface curvature ripples on the electronic properties of monolayer graphene is primarily described via gauge potentials \cite{6}. However, the effect of geometry induced potential {\it \`a la } da Costa \cite{9}  in monolayer graphene has not been investigated. The effect of geometric potential for bilayer graphene is discussed in Ref. [\onlinecite{Avadh}] but in this case the quantum problem is not {\it relativistic}. Indeed, a rigorous derivation of the effect curvature ripples have on the spectrum of the surface carriers, electrons and holes, confined on monolayer graphene  requires a consistent confining procedure for massless Dirac fermions onto a surface embedded in three dimensional space, which we present here.

Low dimensional systems are realized by constraints (holonomic or nonholonomic) that classically reduce the degrees of freedom available for the particle, namely in two-dimensional electron (or hole) systems (2DES), typically realized in III-V semiconductor heterojunctions \cite{7}. The barrier potential at the heterojunction interface, created by differing electron affinities, leads to a clear separation of energy scales for the motion in the 2D plane and motion along the direction normal to the plane. Thus the particle's wave function is separable into normal and surface components and, at low energies, dominated by the surface component \cite{8}. Constraints can thus be realized by introducing external potentials forcing the system to occupy less degrees of freedom \cite{Jensen}.

The quantum properties of a {\it nonrelativistic} particle constrained to an arbitrary orientable surface are known \cite{9}: the geometry of the surface induces an attractive geometric potential in the Schr\"odinger equation in curvilinear surface coordinates
\begin{equation}
V_{\rm da Costa}(q_1,q_2)=-\frac{\hbar^2}{8m}(\kappa_1 - \kappa_2)^2,
\end{equation}
where $m$ is the effective mass of the particle, $\hbar$ is Planck's constant, $(q_1, q_2)$ denote surface coordinates, and $\kappa_1, \kappa_2$ are the two position-dependent principal curvatures of the surface. This potential is purely a result of particle confinement, and is independent of the electric charge of the particle; it is therefore the same for electrons and holes. This result is applicable in the limit
\begin{equation}
\epsilon_0 \kappa \to 0 , 
\end{equation} 
where $\epsilon_0$ is the thickness of the Òtwo-dimensionalÓ surface and 
\begin{equation}\label{eq:kappa}
\kappa={\rm max} \{\kappa_1, \kappa_2  \} .
\end{equation}
is the surface curvature. Here $\epsilon_0$ will correspond to the width of the normal to the surface quantum well in 2DES where particles are confined. However, absent a truly two dimensional system that can be easily bent, the effect of geometric potential on the electronic band structure has been justifiably ignored in device engineering up to now. 
Graphene represents a unique material which can exhibit the effects produced by the  
geometric potential.

Curving graphene can have three major microscopic effects:  (i) changing the distance between carbon atoms, which modifies the hopping amplitude between neighboring  sites and is, in general, very costly due to the large spring constant of graphene $\approx 57$ eV/\AA \cite{spring};  (ii) a rotation of the $p_z$ orbitals and (iii) rehybridization between $\pi$ and $\sigma$ orbitals \cite{epl}. Curvature in  $\rm sp^2$ hybridized carbon allotropes also modifies the Raman spectra through exerting strain in the underlying lattice\cite{raman}. The rotation between the orbitals can be understood within the Slater-Koster formalism \cite{epl, harrison}. When $\pi$ orbitals are not parallel the hybridization between them depends on their relative orientation, which is a function of local curvature. Furthermore, rotation leads to rehybridization between $\pi$ and $\sigma$ orbitals which shifts the energy even further. As a result, Dirac fermions are scattered by ripples and curved regions through a geometric potential $V_G$ which is derived in Section  \ref{sec:constaining}. Section \ref{sec:p-n} discusses the geometric mechanism which shifts the Fermi energy thus leading to the emergence of p-n junctions in monolayer graphene akin to the case of bilayer graphene \cite{Avadh}. Section \ref{sec:bands} deals with periodic corrugations which produce an energy band spectrum. In Section \ref{sec:inverse} we pose the inverse problem, that is how to engineer surfaces with prescribed quantum behavior. In Section \ref{sec:membrane} the membrane aspect of graphene is discussed in view of the constrained two dimensional Dirac equation derived in Section \ref{sec:constaining}.  In Section VII we summarize our main findings.

\section{$\rm p$-$\rm n$ Junctions}\label{sec:p-n}

In this paper, we argue that due to its essentially two-dimensional nature $\epsilon_0 \kappa \to 0,$ gapless linear band structure, and exceptional material strength, monolayer graphene presents an excellent candidate for exploration of two-dimensional systems with curvature. Dimensional analysis implies that for  graphene \cite{Avadh}, the geometric potential $V_G$ must scale as
\begin{equation}
V_G \sim -\hbar v_F |\kappa|
\end{equation}
where $v_F \approx c/300$ is the Fermi velocity of massless carriers in graphene \cite{10}. Here $c$ is the velocity of light in vacuum.

Now, let us discuss the applicability of a constraining procedure to monolayer graphene. To map a curved monolayer graphene sheet on a relativistic quantum problem for a 
particle on a curved surface, two criteria must be satisfied \cite{Avadh}:
\begin{equation}
k_F a_G \ll 1
\end{equation}
and 
\begin{equation}
\kappa a_G \ll 1, 
\end{equation}
where $a_G=1.42 \AA$  is the carbon-carbon distance in graphene, $k_F$ is the Fermi wave vector, and $\kappa$ is the surface curvature given in (\ref{eq:kappa}). The first criterion ensures that monolayer graphene can be treated in the continuum limit and the detailed lattice structure can be ignored. For monolayer graphene near the Dirac point where the relativistic description holds $k_F=0$  and the first criterion is always satisfied.  The second criterion ensures that the detailed lattice structure is not perturbed on the curvature length-scale and the relativistic description holds true because the hexagonal symmetry is not disrupted; and because the thickness of monolayer graphene sheet is $\epsilon_0 \approx 3 a_G$, it also ensures that  $\epsilon_0 \kappa \ll 1 ,$ or that the monolayer graphene sheet with nonzero thickness can be considered a surface. Indeed, for carbon nanotubes with typical radii  $1/\kappa \approx 10-30$ $\AA$ and for typical surface ripples $1/\kappa \approx 100$ $\AA$  on monolayer graphene \cite{11}, the second criterion is satisfied too.

\begin{figure}[ht]
\begin{center}
\includegraphics[scale=0.45]{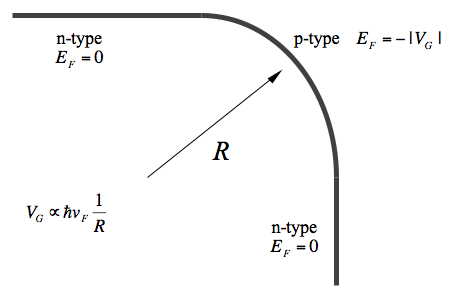}
\caption{\label{p-n} n-p-n junction (transistor) formed by bending monolayer graphene shown in cross-section. The Fermi energy is a function of the position and therefore of curvature $E_F(x)=-|V_G| \propto -\hbar v_F \frac{1}{R}$ along the sheet.}
\end{center}
\end{figure}

As a result, for monolayer graphene $\kappa \gg k_F=0$ the geometric potential $V_G$  becomes dominant over the Fermi energy $E_F.$ For example, let us consider
graphene with n-type carriers in the flat region where $E_F=0,$ then the geometric potential will lower the Fermi energy from the conduction band to the valence band, $E_F=-|V_G| .$ Note that as long as the geometric potential is constant or slowly varying, as is the case for ripples, this result is robust and is independent of any quantum numbers the carriers may have (chirality and/or electric charge). Therefore, the curved regions will have p-type carriers with a natural p-n junction created between the curved and flat regions. For the surface $z=f(x),$ the geometric
potential will create p- and n-type strips where the location, width, and polarity of each strip is determined by the local curvature dependent Fermi energy  $E_F(x),$ see Fig. \ref{p-n} and the derivation which follows.

\section{Constraining the Dirac Equation}\label{sec:constaining}

Let us consider a corrugated graphene sheet along one of the coordinates, for example $x.$ We will demonstrate the confining procedure leading to the appearance of a curvature dependent potential in the Dirac equation describing the carriers, electrons and holes, in graphene. The confining procedure starts with the introduction of an external to the graphene sheet confining potential $V_{\bot}$ which forces the quantum system to remain on the two dimensional surface thus simulating a mechanical constraint \cite{9}. This potential for graphene stems from the Coulomb interaction since as carriers leave the surface of the graphene sheet (their wavefunction is not strictly two dimensional) they distort the charge balance and a potential arises which forces them back on the surface. This potential is similar to the potential with which a charged plate acts on a single charge above it. The properties of $V_{\bot}$ will be introduced later.

The curved two dimensional graphene surface embedded in a flat euclidean three dimensional space is parametrized as follows
\begin{equation}\label{eq:r}
\vec{r}(x,y)=x \vec{e}_x+y \vec{e}_y+f(x) \vec{e}_z,
\end{equation}
where $(\vec{e}_x,\vec{e}_y, \vec{e}_z)$ is the cartesian coordinate system associated with the embedding space, that is the laboratory frame.

In order to derive a moving coordinate system $(\vec{e}_1,\vec{e}_2, \vec{e}_3)$ associated with the surface at each point we need to differentiate the parametric equation  (\ref{eq:r})
\begin{eqnarray}\label{eq:e1}
\vec{e}_1 &&= \left| \frac{\partial \vec{r} }{\partial x}  \right|^{-1} \frac{\partial \vec{r} }{\partial x} = \frac{1}{\sqrt{1+f'^2}}\vec{e}_x+\frac{f'}{\sqrt{1+f'^2}}\vec{e}_z  ,\\\label{eq:e2}
\vec{e}_2  &&= \frac{\partial \vec{r} }{\partial y}= \vec{e}_y ,\\\label{eq:e3}
 \vec{e}_3    &&=  \left| \frac{\partial \vec{r} }{\partial x}  \right|^{-1} \frac{\partial \vec{r} }{\partial x} \times \frac{\partial  \vec{r} }{\partial y} \\
 \nonumber &&= -\frac{f'}{\sqrt{1+f'^2}}\vec{e}_x + \frac{1}{\sqrt{1+f'^2}} \vec{e}_z  .
\end{eqnarray}
The vector $ \vec{e}_3$ is the normal to the surface at each point. The other two vectors $ \vec{e}_1$ and $ \vec{e}_2$ span the tangent space at each point. 
Hereafter $f'=df/dx$ denotes differentiation with respect to $x.$ The moving frame $(\vec{e}_1,\vec{e}_2, \vec{e}_3)$ can be related to the immobile laboratory one
\begin{equation}\label{eq:ex<->ej}
\left(\begin{array}{c}\vec{e}_x \\\vec{e}_y \\ \vec{e}_z\end{array}\right) = T \left(\begin{array}{c}\vec{e}_1 \\\vec{e}_2 \\ \vec{e}_3\end{array}\right), \quad T^{-1}=T^{t}  ,
\end{equation}
where
\begin{equation}\label{eq:T}
T=\left(\begin{array}{ccc}\frac{1}{\sqrt{1+f'^2}} & 0 & -\frac{f'}{\sqrt{1+f'^2}} \\0 & 1 & 0 \\\frac{f'}{\sqrt{1+f'^2}} & 0 & \frac{1}{\sqrt{1+f'^2}}\end{array}\right) .
\end{equation}
From the above two formulas we establish two important relations to be used later in the calculations
\begin{eqnarray}\label{eq:exe1e3}
\vec{e}_x &=& \frac{1}{\sqrt{1+f'^2}}\vec{e}_1 - \frac{f'}{\sqrt{1+f'^2}} \vec{e}_3 ,\\\label{eq:eze1e3}
\vec{e}_z &=& \frac{f'}{\sqrt{1+f'^2}}\vec{e}_1 + \frac{1}{\sqrt{1+f'^2}} \vec{e}_3 .
\end{eqnarray}
Next we introduce a coordinate system of the embedding space associated with the corrugated graphene sheet (\ref{eq:r})
\begin{equation}\label{eq:R}
\vec{R}(x,y,\epsilon) = \vec{r}(x,y) + \epsilon \vec{e}_3 (x).
\end{equation}
Here $\epsilon$ measures the deviation from the surface in normal direction along $\vec{e}_3.$ The metric tensor of the embedding space can be computed according to
\begin{equation}
G_{ij}=\frac{\partial \vec{R}} {\partial i}. \frac{\partial \vec{R}} {\partial j},
\end{equation}
where
\begin{eqnarray}
\frac{\partial \vec{R}} {\partial x}&&= \sqrt{1+f'^2} \vec{e}_1 + \epsilon  \frac{\partial \vec{e}_3} {\partial x } ,\\
\frac{\partial \vec{R}} {\partial y}&&=  \vec{e}_2 ,\qquad
\frac{\partial \vec{R}} {\partial \epsilon}= \vec{e}_3 .
\end{eqnarray}
A straightforward computation yields
\begin{equation}
\frac{\partial \vec{e}_1} {\partial x} = - \kappa \vec{e}_3, \qquad\frac{\partial \vec{e}_3} {\partial x} =  \kappa \vec{e}_1, 
\end{equation}
where hereafter the nonvanishing curvature $\kappa$ of the graphene sheet is
\begin{equation}\label{eq:kappa_f}
\kappa=- \frac{f''}{1+f'^2},
\end{equation}
and should not be mistaken for (\ref{eq:kappa}).

In the mobile basis $(\vec{e}_1,\vec{e}_2, \vec{e}_3)$ the metric tensor is
\begin{equation}
G=\left(\begin{array}{ccc} \Delta^2 & 0 & 0 \\0 & 1 & 0 \\0 & 0 & 1\end{array}\right),
\end{equation}
where 
\begin{equation}\label{eq:Delta}
\Delta=\sqrt{1+f'^2} + \epsilon \kappa.
\end{equation}

In the moving orthonormal frame the gradient operator becomes
\begin{equation}
\vec{\nabla}=\vec{e}_1 \frac{1}{\Delta} \frac{\partial }{\partial x} + \vec{e}_2 \frac{\partial }{\partial y}  + \vec{e}_3 \frac{\partial }{\partial \epsilon}.
\end{equation}
With units $\hbar = v_F =1$ the hamiltonian of the massless Dirac equation $\mathcal{H}\Psi=E\Psi$ describing carriers, belonging to different sublattices (via the two component spinor wavefunction $\Psi$) in corrugated graphene is $\mathcal{H}=-i\vec{\alpha}. \vec{\nabla}.$ Here $\vec{\alpha}$ is a vector consisting of $2\times2$ Pauli matrices.

In the coordinate system (\ref{eq:R}) the Dirac hamiltonian takes the form
\begin{equation}
\mathcal{H}=-i\vec{\alpha}.\vec{e}_1\frac{1}{\Delta} \frac{\partial }{\partial x} -i\vec{\alpha}.\vec{e}_2 \frac{\partial }{\partial y}  -i\vec{\alpha}. \vec{e}_3 \frac{\partial }{\partial \epsilon} .
\end{equation}

The wave function $\Psi$ is normalized according to 
\begin{equation}
\int   \Psi^{\dag} \Psi \; \Delta dx dy d\epsilon =1,
\end{equation}
where $^{\dag}$ denotes hermitian conjugation. The factor $\Delta$ in the integration measure complicates calculations and obscures hermiticity. We define a new wavefunction
\begin{equation}
\psi=\Delta^{1/2} \Psi
\end{equation}
which has the normalization condition
\begin{equation}
\int  \psi^{\dag} \psi \; dx dy d\epsilon =1.
\end{equation}
The original Dirac equation can be recast in terms of an energy operator $\tilde{H}=\Delta^{1/2} \mathcal{H} \Delta^{-1/2}$ which acts on the wave functions $\psi$
\begin{eqnarray}\label{eq:H_psi}
\nonumber  \tilde{H}=&& -i\vec{\alpha}.\vec{e}_2 \frac{\partial }{\partial y}  -i\vec{\alpha}. \vec{e}_3 \frac{\partial }{\partial \epsilon}+ i \frac{\kappa }{2\Delta}\vec{\alpha}. \vec{e}_3\\
&& -\frac{i}{\Delta}\vec{\alpha}.\vec{e}_1\left(  \frac{\partial }{\partial x} - \frac{1}{2\Delta}\frac{\partial \Delta}{\partial x} \right) . 
\end{eqnarray}
This hamiltonian is difficult to compute with because the matrices $\vec{\alpha}.\vec{e}_j$ are $x$-dependent. The momentum operators here are expressed with respect to the moving coordinate system $\vec{e}_j$ while the pseudo-spin operators $\vec{\alpha}$ are expressed with respect to some fixed rectilinear frame. To remedy this we change the pseudo-spin basis through a unitary transformation $\Omega(x)$
\begin{equation}\label{eq:defOmega}
\Omega(x)\vec{\alpha}.\vec{e}_j  (x) \Omega^{\dagger}(x)=\alpha_j 
\end{equation}
holding true for all $x.$ Here the $x$-dependence of the unit vector $\vec{e}_j$ is made explicit.

To begin with, the moving frame undergoes an evolution as traversing the $x$ direction
\begin{equation}\label{eq:de_j/dx}
\frac{d \vec{e}_j}{dx}=\vec{\omega} \times \vec{e}_j, \quad \rm {for} \quad j=1,2,3. 
\end{equation}
Here 
\begin{equation}
\vec{\omega}=\vec{e}_3 \times \frac{d \vec{e}_3}{dx}= -\kappa \vec{e}_2
\end{equation}
is a smoothly varying instantaneous ``angular velocity" associated  with the evolution of the frame. 

Next, we suppose $\Omega(x)=\mathcal{P}\exp{\left[ \frac{i}{2}\int^x dx' \vec{\Sigma}.\vec{\omega}(x')   \right]},$ where $\mathcal{P}$ denotes path ordering along $x$ and $\vec{\Sigma}$ is to be determined. This yields the following equation for $\Omega$
\begin{equation}\label{eq:dOmega/dx}
\frac{d \Omega}{dx}=\frac{i}{2}\vec{\Sigma}.\vec{\omega}\Omega.
\end{equation} 
After differentiating (\ref{eq:defOmega}) with respect to $x,$ we arrive at
\begin{equation}
\frac{d \Omega}{dx}\vec{\alpha}.\vec{e}_j  \Omega^{\dagger} + \Omega\vec{\alpha}.\vec{e}_j  \frac{d \Omega^{\dag}}{dx} + \Omega\vec{\alpha}. \frac{d \vec{e}_j}{dx} \Omega^{\dagger}  = 0.
\end{equation}
 Now using (\ref{eq:de_j/dx}) and (\ref{eq:dOmega/dx}) we obtain
\begin{equation}\label{eq:sigma}
\left[ \vec{\Sigma}.\vec{\omega}, \alpha_j \right] = 2 i \Omega\vec{\alpha}. \frac{d \vec{e}_j}{dx} \Omega^{\dagger}.
\end{equation}
The form we have chosen for $\Omega$ points to a change of pseudo-spin basis to spinors representing carriers in sublattice A and B  (Fig. 1) in graphene's corrugated surface described locally by $\vec{e}_1$ and $\vec{e}_2 .$
 
We write down a small computation table
\begin{eqnarray*}
\left[ \vec{\Sigma}.\vec{\omega}, \alpha_1 \right] &=& - 2 i \kappa \alpha_3 ,\\
\left[ \vec{\Sigma}.\vec{\omega}, \alpha_2 \right] &=& 0 ,\\
\left[ \vec{\Sigma}.\vec{\omega}, \alpha_3 \right] &=& 2 i \kappa \alpha_1 .
\end{eqnarray*}
From the middle equation follows $\vec{\Sigma}.\vec{\omega}= C \alpha_2.$ If we choose $C=\kappa$ the above table transforms into
\begin{eqnarray*}
\left[ \alpha_2, \alpha_1 \right] &=& -2 i \alpha_3 ,\\
\left[\alpha_2, \alpha_2 \right] &=& 0 ,\\
\left[ \alpha_2, \alpha_3 \right] &=& 2 i \alpha_1 .
\end{eqnarray*}
These relations coincide with the algebra of the Pauli matrices $[\sigma_a, \sigma_b] = 2 i \varepsilon_{abc} \sigma_c \Rightarrow \sigma_1 \sigma_2=i \sigma_3.$

We define a new spinor wavefunction to take advantage of this transformation
\begin{equation}
\chi(x,y,\epsilon)=\Omega(x)\psi(x,y,\epsilon)
\end{equation}
in addition to  a new Dirac hamiltonian $H=\Omega \tilde{H} \Omega^{\dag}$ acting on $\chi.$ Although this transformation replaces the $\alpha$-matrices as intended, it generates a gauge term in $H$: 
\begin{equation}
-\frac{i}{\Delta} \Omega \vec{\alpha}.\vec{e}_1 \frac{d \Omega^{\dag}}{dx} = - \frac{1}{2\Delta} \alpha_1  \vec{\Sigma}.\vec{\omega}.
\end{equation}
Using the above given tables we find 
\[
 \alpha_1  \vec{\Sigma}.\vec{\omega}=i\kappa \alpha_3 \Rightarrow -\frac{i}{\Delta} \Omega \vec{\alpha}.\vec{e}_1 \frac{d \Omega^{\dag}}{dx} =  -\frac{i\kappa}{2\Delta} \alpha_3.
\]
This term cancels the third term in (\ref{eq:H_psi}). Finally, the three dimensional Dirac hamiltonian is simplified

\begin{eqnarray}\label{eq:H}
\nonumber H && = -i\alpha_2 \frac{\partial }{\partial y}  -i\alpha_3 \frac{\partial }{\partial \epsilon}\\
&& -\frac{i}{\Delta} \alpha_1\left( \frac{\partial }{\partial x}  - \frac{1}{2\Delta}\frac{\partial \Delta}{\partial x} \right) . 
\end{eqnarray}

\begin{figure}[ht]
\begin{center}
\includegraphics[scale=0.4]{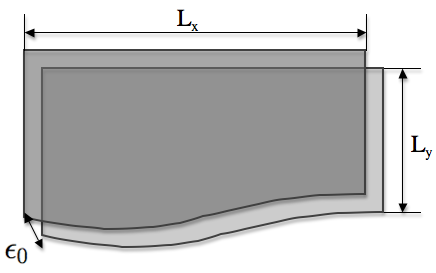}
\caption{\label{fig:constraining} The gradual confining procedure $\epsilon_0 \to 0$ and the domain where the Dirichlet boundary conditions apply for (\ref{eq:H}).}
\end{center}
\end{figure}

The transformations we have performed reduce the original Dirac problem to the study of this perturbed Dirac hamiltonian in the space $-\epsilon_0 /2 \le \epsilon  \le \epsilon_0 /2,$ where $\epsilon_0 \to 0,$ between two parallel identical corrugated surfaces $ 0 \le x \le L_x,$ $0 \le y \le L_y$ with respect to a fixed rectilinear frame (see Fig. \ref{fig:constraining}). This simplification gave rise to a variety of curvature dependent terms in (\ref{eq:H}).

Despite its two dimensional nature, graphene has three acoustic phonon modes \cite{phonon}. The two in-plane modes (LA, TA) have a linear dispersion relation, whereas the out of plane mode (ZA) has a quadratic dispersion relation pointing to the possibility of having an effective mass term $\alpha_1 m_{\bot}$ included in the above hamiltonian which is three dimensional and contains information for the off-surface processes as well. As soon as the carriers leave the two dimensional world of the graphene sheet where symmetry precludes the mass, they attain mass. Therefore we will treat the transverse dynamics in the  non-relativistic limit. Furthermore, according to the ideas presented in the beginning of this section, we shall now consider the idealistic spatial potential $V_{\bot}(\lambda_0, \epsilon),$ where $\lambda_0$ measures the squeezing strength of this potential
\begin{equation}
\lim_{\lambda_0 \to \infty} V_{\bot}(\lambda_0, \epsilon) = \left\{\begin{array}{ll}0, & \epsilon=0, \\\infty, & \epsilon \neq 0. \end{array}\right. 
\end{equation}
For example, we can imagine a realistic harmonic binding $V_{\bot}(\lambda_0, \epsilon)= \frac12 m_{\bot}\lambda_0^2 \epsilon^2$ with $\lambda_0$ going to infinity which yields $\langle \epsilon^2 \rangle \approx \hbar/m_{\bot}\lambda_0$ in the nonrelativistic limit.

Next we look for eigenstates of (\ref{eq:H}) in  the form
\begin{equation}\label{eq:sum_ansatz}
\chi(x,y,\epsilon)=\sum_{\lambda} \psi_{\lambda} (x,y) h_{\lambda} (\epsilon),
\end{equation}
where $h_{\lambda} (\epsilon)$ are eigenstates of the transverse one dimensional hamiltonian
\begin{equation}\label{eq:H_perp}
H_{\bot}= -i\alpha_3 \frac{\partial }{\partial \epsilon} + \alpha_1 m_{\bot} + V_{\bot}(\epsilon).
\end{equation}
The label $\lambda$ denotes the quantum numbers necessary to specify the eigenstates $h_{\lambda}(\epsilon)$ of $H_{\bot}$ completely and physically speaking it is the ``transverse energy'' due to confinement. The boundary conditions are vanishing (Dirichlet type) at $\epsilon=\pm \epsilon_{0}/2.$ We need to take the nonrelativistic limit when solving $H_{\bot}$ in order to overcome the well known jittery behavior (due to Heisenberg principle) of the wavefunction when confined in too small an interval $\epsilon_0$ \cite{confinement}.

The remaining two dimensional part amounts to
\begin{equation}\label{H:lim}
H'= -i\alpha_2 \frac{\partial }{\partial y} -\frac{i}{\Delta} \alpha_1\left( \frac{\partial }{\partial x}  - \frac{1}{2\Delta}\frac{\partial \Delta}{\partial x} \right).
\end{equation}
Since we have already factored out the transverse  part, we take the limit $\epsilon \to 0$ in (\ref{H:lim}). This action is in agreement with the standard procedure in constrained quantum dynamics in the case of small $\epsilon_0 \approx 3 a_G$\cite{9}. The smallness of  $\epsilon_0$ in graphene is combined with the relatively small curvature of the naturally occurring corrugations. Thus for graphene the following applies
\begin{equation}
|\epsilon\kappa| \ll 1,  \qquad  \left|\epsilon\frac{\partial \kappa}{\partial x}\right| \ll 1, 
\end{equation}
rendering the approximation
\begin{equation}
\Delta \approx \sqrt{1+f'^2   }
\end{equation}
justified. Indeed,
\[
H'= -i\alpha_2 \frac{\partial }{\partial y} -i\alpha_1\left(  \frac{1}{\sqrt{1+f'^2   }}\frac{\partial }{\partial x}  + \frac{\kappa(x)}{2}\frac{f'}{ \sqrt{1+f'^2   }} \right).
\]
Introducing a new variable, which is the line length along the corrugated surface  
\begin{equation}
s=\int_{0}^{x} \sqrt{ 1 + f'^2  }dx',
\end{equation}
instead of $x$ and
\begin{equation}
K[s(x)]= \kappa(x) \frac{f'}{ \sqrt{1+f'^2   }},
\end{equation}
we may rewrite $H'$ as
\begin{equation}\label{eq:H2d}
H'= -i\alpha_2 \frac{\partial }{\partial y} -i\alpha_1\left( \frac{\partial }{\partial s}  - V_G(s) \right),
\end{equation}
where
\begin{equation}\label{eq:V_G}
V_G(s)=-\frac{1}{2}K(s) .
\end{equation}

As is often the standard approach with the Dirac equation, it is easier to work with the square of the Dirac operator instead of the first order form $H'^2\psi_{\lambda}=(E-\lambda)^2\psi_{\lambda}=E_{el}^2 \psi_{\lambda},$ where $E$ is the total energy and $E_{el}$ is the energy of the electrons on the surface. We have 
\begin{eqnarray*}
\ H'^2 = -\frac{\partial^2 }{\partial y^2} - \frac{\partial^2 }{\partial s^2}  - \frac{1}{2}\frac{\partial K}{\partial s} -\frac{1}{2} K \frac{\partial }{\partial s}  - \frac{1}{4}K^2(s).
\end{eqnarray*}
Looking for solutions obeying the ansatz
\begin{equation}
\psi_{\lambda} = e^{-\frac14 \int_{0}^{s} K(s')ds' }\eta_{\lambda}
\end{equation}
we simplify the squared hamiltonian acting on $\eta_{\lambda}$  considerably
\begin{eqnarray}\label{eq:H2}
H'^2 = -\frac{\partial^2 }{\partial y^2} - \frac{\partial^2 }{\partial s^2}  + V(s) ,
\end{eqnarray}
where 
\begin{equation}\label{eq:V}
V(s) = - \frac{1}{16} \left[ 4 \frac{\partial K}{\partial s}  + 3 K^2(s) \right].
\end{equation}

The hamiltonian (\ref{eq:H2}) with a non-vanishing potential (\ref{eq:V}) possesses the usual phenomenology of bound and scattering states. Moreover, band structure is allowed for periodic $V(s).$ Remembering the geometrical origin of (\ref{eq:V}), we may state that all of the above mentioned instances for graphene are of geometrical origin including decoherence as a function of scattering and interaction through $V_G$.

\section{Energy Bands}\label{sec:bands}

\begin{figure}[ht]
\begin{center}
\includegraphics[scale=0.4]{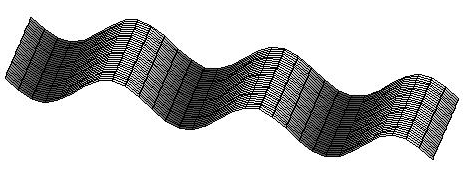}
\caption{\label{periodic} A periodically corrugated surface $f(x)=\sin(x)$.}
\end{center}
\end{figure}

Let us explore a periodically corrugated graphene sheet (see Fig. \ref{periodic} ) parametrized as follows
\begin{equation}
f(x)=\varepsilon \sin\left(\frac{x}{a}\right). 
\end{equation}
Suppose $({\varepsilon}/{a})^2 \ll 1$ is a small parameter then the following approximation holds true up to second order in this small parameter $1+f'^2 \approx 1$ and $s \approx x.$

For the potential $V(s)$ we obtain
\[
V(x)=-\frac{1}{4} \frac{\varepsilon^2}{a^4} \cos \left( \frac{2x}{a}  \right) - \frac{3}{2^7} \frac{\varepsilon^4}{a^6} \approx  -\frac{1}{4} \frac{\varepsilon^2}{a^4} \cos \left( \frac{2x}{a}  \right).
\]
Introducing the dimensionless variable $z=s/a,$  the $s$-component of the squared Dirac hamiltonian (\ref{eq:H2}) acting on a factorized wave function $\eta_\lambda (y,s)=\theta(y) \phi(s)$ takes the form of the Mathieu equation
\begin{equation}
\left[\frac{d^2}{dz^2}  + \alpha -2 q  \cos \left( 2 z  \right) \right] \phi(z)=0,
\end{equation}
where $\alpha=a k_s$ and $q=-\frac18  \frac{\varepsilon^2}{a^2}.$ Here $k_s$ is the wave number along the $s$-degree of freedom. The solutions of the above are given in terms of Mathieu functions \cite{Abram&Steg} and a band gap structure naturally emerges as a consequence of the properties of the Mathieu equation \cite{Abram&Steg}. The $y$-component of the solution obeys the standard harmonic oscillator equation and will not be discussed here.

According to the Floquet's theorem, the general solution to the Mathieu equation is
\begin{equation}
\phi(z)=c_1 e^{\xi z} h(z) +
c_2 e^{-\xi z} h(-z),
\end{equation}
where $h(z)$ is a periodic complex valued function with period $2 \pi$ and the characteristic  exponent $\xi$ is a definite complex valued function of $\alpha$ and $q.$ Here the constants $c_i$ are arbitrary up to $c_1 c_2 \neq 0$. The characteristic  exponent $\xi=i {\rm p}$ is imaginary for  spatially undamped solutions, while it is real or complex for  spatially damped solutions\cite{Brillouin}. Spatially undamped solutions exist in specific regions only defined by $\xi=\xi[\alpha(a, q_s),q(a, \epsilon)]$ and are refered to as {\it allowed energy bands.} There are four types of periodic solutions, even $z$-parity $ce_{\rm p}(q, z)$ and odd $z$-parity $se_{\rm p}(q, z)$ in two classes, even-integer at ${\rm p} =2 {\rm n}$ and odd-integer at ${\rm p} =2 {\rm n} + 1,$ with respective eigenvalues (respectively eigenenergies) $a_{\rm p}(q)$ to $ce_{\rm p}(q, z)$ and $b_{\rm p}(q)$ to $se_{\rm p}(q, z).$ The functions are normalized so as $(1/ \pi) \int_0^{2\pi}   [\phi(z)]^2 d z=1.$ At small $|q|=\left| -\frac18  \frac{\varepsilon^2}{a^2}\right| \ll 1,$ $ce_{\rm 0}(q, z) \approx 1,$
\begin{equation}
ce_{\rm p}(q, z) \approx  \cos{({\rm p} z)}, \quad
se_{\rm p}(q, z) \approx  \sin{({\rm p} z)}
\end{equation}
and $a_{\rm p} \approx  {\rm p}^2,$ $b_{\rm p} \approx  {\rm
p}^2.$ These integer class periodic functions define the edges of
the allowed energy bands.

Mathieu eigenvalue equation defines allowed energy bands at $q<0$ (upper-sign branch) and at $q>0$ (lower-sign branch) respectively\cite{Brillouin, Abram&Steg}. As $q$ is increased, the allowed bands get narrower turning to single levels which is not the case here since $q<0$.
The allowed band edges in the energy versus $q$ plane are described by Mathieu eigenvalues $a_{\rm p}(q)$ and $b_{\rm p}(q),$ or alternatively, by the corresponding periodic eigenfunctions $ce_{\rm p}(q, z)$ and $se_{\rm p}(q, z)$ at integer ${\rm p}$  $({\rm p}=1,2,3,\ldots),$ while the solution for each energy in the interior of the band is generated by pairs of Mathieu functions at intermediate non-integer $\rm p.$  Whithin an allowed band, Mathieu eigenvalues and eigenstates are continuous  functions of the wave number $k={\rm p} \pi,$ where $0 \leq {\rm p} \leq 1$ for the first Brillouin zone\cite{Brillouin}. At $q<0,$  which is the case here, the allowed bands are defined as follows: $(a_{\rm 0}, a_{\rm 1}), (b_{\rm 1}, b_{\rm 2}), (a_{\rm 2}, a_{\rm 3}),$ etc.

\section{Inverse Problem}\label{sec:inverse}

We showed that the relativistic quantum dynamics of massless carriers in corrugated graphene can be mapped on a two dimensional Schr\"odinger problem (\ref{eq:H2}) with the geometric potential (\ref{eq:V}). Now we pose the inverse problem (for the nonrelativistic case it is discussed in Ref. [\onlinecite{victor}]), namely we may impose a prescribed potential $U(s)$ and solve for the graphene surface profile $f[s(x)]$, i.e. we solve the following third order nonlinear ordinary differential equation in terms of $f(s)$ given some boundary conditions
\begin{equation}
U[s(x)]=V[s(x)].
\end{equation} 
However, here we will focus on a simpler problem. For the constrained two dimensional Dirac equation, where the geometric potential is
$
V_G[s(x)],
$
we can pose a similar inverse problem for a prescribed potential $R(s)$
\begin{equation}
R(s)=V_G(s).
\end{equation}
Let us answer the question which surfaces do \textit{not} produce a geometric potential, that is
\begin{equation}
R(s)=0 \Rightarrow V_G[s(x)]=0.
\end{equation}
Using (\ref{eq:V_G}) we can show that this is equivalent to
\begin{equation}
f'' f' =\frac12 \frac{d}{dx} f'^2 = 0
\end{equation}
The solution for the graphene profiles which do not produce geometric potential corresponds to  \textit{flat} surfaces only
\begin{equation}
f(x)=C_1 x + C_2,
\end{equation}
where $C_1$ and $C_2$ are arbitrary real constants. Any corrugation will generate a geometric potential $V_G \neq 0.$

\section{An Electronic Membrane}\label{sec:membrane}

Graphene is an example of an electronic membrane apt to gating and probing \cite{epl} which adds a new perspective to the physics of membranes \cite{nelson}. In this section we establish a connection between the membrane aspect and the electronic properties of graphene through the geometry induced potential $V_G$ and the Dirac equation constrained in two dimensions (\ref{eq:H2d}). To build an effective theory we introduce the bending free energy $E_{bend}$ describing an almost flat membrane in the thermodynamic limit \cite{thd}
\begin{equation}
E_{bend}=\int d q_1 d q_2 \sqrt{g} \left[  \gamma + \frac12 \mu (2 M)^2 + \nu \mathcal{K}  \right],
\end{equation}
where $(q_1, q_2)$ are the surface's coordinates, $g$ is the determinant of the metric, $M$ and $\mathcal{K}$ are the Mean and the Gaussian curvature, respectively. Here $\gamma$ is the tension, $\mu$ is the bending rigidity and $\nu$ is the Gaussian rigidity. Besides $\mu \approx 1$ eV \cite{bendrig}, $\gamma$ and $\nu$ are presently not known. If we neglect the Van der Waals interaction between the graphene membrane and an external support, the resulting free energy using formulas from Refs. [\onlinecite{epl, membrane}] is  
\begin{equation}
E_{bend}=\int d x d y \sqrt{1+f'^2} \left[  \gamma  + \frac12 \mu \kappa^2 \right],
\end{equation}
where $\kappa$ is given by (\ref{eq:kappa_f}). The Gaussian total curvature term vanishes because it is a total derivative and for a membrane with a fixed topology it gives a constant (which is neglected) according to the Gauss-Bonet theorem \cite{membrane2}. Now, the above expression has to be combined with the following energy functional stemming from  (\ref{eq:sum_ansatz}) and (\ref{eq:H2d})
\begin{eqnarray}
E_{el}=&&\sum_\lambda \int dx dy d\epsilon |h_{\lambda}(\epsilon)|^2  \psi_\lambda^\dag \left[  
-i\alpha_2 \frac{\partial }{\partial y} 
 \right. \\
\nonumber && \left.  -i\alpha_1\frac{1}{\sqrt{1+f'^2}}\frac{\partial }{\partial x}  + i\alpha_1 V_G(x)  \right] \psi_\lambda.
\end{eqnarray}
The coupled problem $\delta (E_{bend}+E_{el})= 0$ has to be solved self-consistently in order to give graphene's equilibrium corrugations, which will be a function of the confined eigenstates $h_{\lambda}(\epsilon)$ and the strength of the confining potential $V_{\bot}(\epsilon).$

\section{Conclusion}

In conclusion, we have demonstrated that for n-doped graphene the geometric potential $V_G$ due to confinement in the Dirac equation for the massless carriers dominates over the Fermi energy $E_F$ in the bent regions (where the curvature is nonvanishing) and leads to the formation of alternating p- and n- type regions, that is p-n junctions of pure geometrical origin. This may have a practical application in constructing wavelength-specific (because $V_G$ can be tuned) solar cells based on graphene. Next, we have derived proper constraining procedure leading to the two dimensional Dirac equation for the carriers in corrugated graphene.  The geometric potential due to confinement $V_G$ emerges as a consequence of the quantum mechanical property that the wavefunction is three dimensional and when constrained ``experiences" the curvature of the sheet. 
The inverse problem, namely the engineering of graphene surfaces with prescribed quantum properties, is posed. It opens an avenue for device engineering based on corrugated graphene. Periodic corrugations lead to the emergence of band gap structure in the energy spectrum, which can be tuned by varying the amplitude and the period  of the periodic corrugation. In general, corrugated graphene is a perfect material with respect to geometrical effects due to the scale of the geometric interaction $V_G \sim \hbar v_F |\kappa|,$ where for $\kappa \approx 100^{-1}\;  {\rm nm}^{-1}$ (naturally formed ripples) equals $V_G \approx 6\; \rm meV$ which is measurable. We believe that our results will lead to experimental verification of the phenomena suggested here.

\section{Acknowledgments} 

We acknowledge initial discussions with Y.N. Joglekar.  This work was supported in part by the 
U.S. Department of Energy and in part by project IT-QuantTel, as well as from Funda\c{c}\~{a}o para a Ci\^{e}ncia e a
Tecnologia (Portugal) and FEDER (European Union), namely via project PTDC/EEA-TEL/103402/2008 QuantPrivTel.

\end{document}